\documentstyle[12pt]{article}

\catcode`\@=11

\@addtoreset{equation}{section}
\@addtoreset{equation}{subsection}
\def\theequation{\ifnum\value{subsection}>0\relax
\thesubsection.\arabic{equation}\relax
\else\ifnum\value{section}>0\relax
\thesection.\arabic{equation}\relax
\else\arabic{equation}\fi\fi}

\begin{document}
\begin{titlepage}
\rightline{SNUTP-95-061,EWHA-TH-005}
\rightline{May 1995}
\vskip 1cm
\centerline{\Large\bf  ${\bf gl(n|m)}$ color Calogero-Sutherland models}
\centerline{\Large\bf and Super Yangian Algebra}
\vskip 1cm
\centerline{Changrim Ahn$^{\dagger}$ and Wai Ming Koo$^{\ddagger}$}
\vskip 1cm
\centerline{$^{\dagger}$ Department of Physics}
\centerline{Ewha Womans University, Seoul 120-750, Korea}
\vskip .5cm
\centerline{$^{\ddagger}$Center for Theoretical Physics}
\centerline{Seoul National University, Seoul 151-742, Korea}
\vskip 2cm
\begin{abstract}
A supersymmetric extension of the color Calogero-Sutherland
model is considered based on the Yangian $Y(gl(n|m))$.
The algebraic structure of the model is discussed in some
details. We show that the commuting conserved quantities
can be generated from the super-quantum determinant,
thus establishing the integrability of the model.
In addition, rational limit of the model is studied where
the Yangian symmetry degenerates into a super loop algebra.
\end{abstract}
\end{titlepage}
\newpage

\section{Introduction}

The Calogero-Sutherland model and its $gl(n)$ extension\cite{gln}
(collectively called CS model here) are
integrable models based on a different kind mechanism for their
integrablilty. Unlike integrable models with short range interactions,
the CS model's Hamiltonian and the full set of commuting conserved
quantities can not be obtained from the trace of the $T$-matrix
that gives the generators of the quantum algebra. This
is because the CS model has a Yangian symmetry\cite{cha} even with finite
number of particles and the trace of the $T$-matrix does not
commute with the Yangian algebra generators. Instead it turns out
that the generating function for the commuting conserved quantities
is the quantum determinant\cite{sac}. Obvious as it may seems,
this is not true
however for the case of Haldane-Shastry (HS) spin chain\cite{hs},
which is an integrable spin chain
with long range interaction and a Yangian symmetry. The quantum
determinant in this case is proportional to the identity operator.
So far it is not clear what the generating function for the
commuting conserved quantities of the HS spin chain should be, other
than using the approach of \cite{poly}.

Restricting to the CS model, the algebraic structure of the model
can be summarized as follows\cite{sac,mge}: The model constitutes
an $N$-body quantum
mechanical system; the particles are described by their positions
$x_j\;,j=1\ldots N$ and their $n$-color degrees of freedom labeled by
$a_j$. The configuration space furnishes a representation for the Yangian
$Y(gl(n))$; one starts with $N$ copies of fundamental
representation of $gl(n)$ build from the color indices of the particles,
which can be taken to be the set of matrix unit $X_j^{ab}\;;a,b=1,\ldots,n$.
Then constructs the Yangian generators $T_0^{ab}, T_1^{ab}$ using this
representation of $gl(n)$ and the variables $x_j$ as follows:
\begin{eqnarray}
T^{ab}_0&=&\sum_{j=1}^N X^{ab}_j  \\
T^{ab}_1&=&\sum_{j,k=1}^N X^{ab}_j\left(\delta_{jk}\frac{\partial}
{\partial x_j}+\lambda\theta_{jk}P_{jk}\right)
\end{eqnarray}
where $\theta_{jk}$ is some $\bf C$ number function that
depends only on $x_j-x_k(\neq 0)$, and $P_{jk}$ is an operator
that exchanges the color degrees of freedom of particles $j,k$.
It has the expression
\[P_{jk}=\sum_{a,b=1}^{n}X^{ab}_jX^{ba}_k\]
in terms of the matrix unit.

Requiring that $T^{ab}_0,T^{ab}_1$ satisfy the Yangian relations
encoded in following equation
\begin{equation}
R_{oo'}(u-v)T_0(u)T_{0'}(v)=T_{0'}(v)T_0(u)R_{oo'}(u-v)\label{eq:yan}
\end{equation}
where
\begin{equation}
R_{oo'}(u)=u{\bf 1}+\lambda P_{oo'}\label{eq:rmx}
\end{equation}
with $o,o^{'}$ denote the two $n-$dimensional auxiliary spaces, and
\begin{equation}
T_o(u)=\sum_{a,b=1}^{n}X_o^{ba}\left({\bf 1}\delta_{ab}+\sum_{s\geq 0}u^{-s-1}
T_s^{ab}\right)\;,
\end{equation}
(similarly for $T_{o^{'}}(v)$,) one deduces that $\theta_{jk}$ has to satisfy
the relation
\begin{equation}
\begin{array}{rll}
&\theta_{jk}\theta_{jn}+\theta_{jk}\theta_{nk}-\theta_{jn}\theta_{nk}=
\theta_{jk}\;;\hspace{1cm}&j\neq k\neq n\neq j\\
\mbox{and }\hspace{1cm}&\theta_{jk}+\theta_{kj}=1\;;\hspace{3cm}&j\neq k\;.
\end{array} \label{eq:fre}
\end{equation}
The rest of the Yangian generators $T_s^{ab}\;,s>1$ can be generated
from $T_0^{ab}, T_1^{ab}$ using the defining relation (\ref{eq:yan}),
one can then compute the quantum determinant defined as
\begin{equation}
\mbox{qdet}T(u)=\sum_{\sigma\in{\bf S}_s}
(-1)^{\mbox{\footnotesize sign}(\sigma)}
T^{1\sigma(1)}(u)T^{2\sigma(2)}(u+\lambda)\cdots T^{n\sigma(n)}(u+(n-1)\lambda)
\;.
\end{equation}
The quantum determinant generates the center of the Yangian algebra
and expanding it as formal power series in $u^{-1}$ as
\begin{equation}
\mbox{qdet}T(u)={\bf 1}+u^{-1}Q_0+u^{-2}Q_1+u^{-3}Q_2+\cdots\;,\label{eq:det}
\end{equation}
the operators $Q_j$'s commute with the Yangian generators and therefore
among themselves. Moreover, the Hamiltonian
\begin{equation}
H=\sum_{j,k=1}^{N}\left(\delta_{jk}\frac{\partial^2}{\partial x_j^2}
+\frac{\lambda(\lambda-P_{jk})}{\sin^2 (x_j-x_k)}\right)
\end{equation}
can be obtained from $Q_2$. So the rest of the operators $Q_j$'s are
conserved quantities. In particular, $Q_0$ and $Q_1$ contain the
particle number operator (which is $N$ here) and the total momentum
operator.

The aim of this paper is to show that one can extend the $gl(n)$
color CS model to the $gl(n|m)$ case, i.e. the particles now carry color
indices which
form a representation of the $gl(n|m)$ graded Lie algebra. The model in
the simplest case where $n=m=1$ and in the 'rational limit' (see later) is the
model studied in \cite{men} under the name supersymmetric CS model without the
harmonic potential, for this reason we also call this $gl(n|m)$ extension
the supersymmetric CS model even though in this case the Hamiltonian can not
be written as a commutator of some supersymmetric charges $Q,Q^{\dagger}$
with the property $(Q^{(\dagger)})^2=0$. In general, the supersymmetric
CS model has a Yangian symmetry $Y(gl(n|m))$, which is the Yangian
deformation of the universal enveloping algebra of the
super (graded) Lie algebra $gl(n|m)$, and
algebraic structure described above still prevails in this extension.
We give an explicit
construction of the representation of the supersymmetric Yangian generators
based on this model and show that the Hamiltonian has exactly the same
form as that of the CS model, except that the exchange operator now
is built out of fundamental representation of the $gl(n|m)$ generator which
has the property that when exchanging two fermionic colors an additional
factor $(-1)$ is produced. The generator of the
commuting conserved quantities in this
case is the supersymmetric generalization of the quantum determinant.

The construction of the SUSY CS model is given in detail in section 3 after
an elementary reminder of the Yangian $Y(gl(n|m))$ in
section 2. We then consider in section 4 various limiting cases of the
model, which includes the rational limit where the Yangian symmetry
degenerates into the loop algebra symmetry. Section 5 contains discussion
and conclusion.

\section{The Supersymmetric Yangian}

Consider the graded Lie algebra $gl(n|m)$ defined by
the set of generators $e^{ab}\;,a,b=1,\cdots,n+m$ that satisfy
the following relation
\begin{equation}
\left[e^{ab}\right.,\left.e^{cd}\right\}=
\delta_{bc}e^{ad}-(-1)^{(p(a)+p(b))(p(c)+p(d))}
\delta_{ad}e^{cb}
\end{equation}
where $p(a)$ is the ${\bf Z}_2$ grading defined as
\[p(a)=\left\{\begin{array}{ll}
           0&\hspace{2cm}a=1,\cdots,n\\
           1&\hspace{2cm}a=n+1,\cdots,n+m
\end{array}\right.\]
and $[\;\;,\;\;\}$ is the graded Lie bracket.

Let  $V$ be an $n+m$ dimensional $Z_2$ graded vector space and
$\{v^a\;,a=1,\cdots,n+m\}$ be a homogeneous basis whose grading is given
as above.   The matrix unit $X^{ab}$ defined as
\begin{equation}
X^{ab}v^c=\delta_{bc}v^a
\end{equation}
furnishes a vector representation $\rho$ for $gl(n|m)$ as follows
\begin{eqnarray}
\rho(e^{ab})&=&X^{ab}\\
\left[X^{ab}\right.,\left.X^{cd}\right\}&
\equiv&X^{ab}X^{cd}-(-1)^{(p(a)+p(b))(p(c)+p(d))}
X^{cd}X^{ab} \label{eq:lieb}
\end{eqnarray}
where the second line defines the graded bracket.

In the sequel, we shall refer to the vector $v^a$ with
grading $p(a)=0(1)$ as the bosonic- (fermionic-) color state, and
the term supersymmetric arises from the fact that
element $e^{ba}$ with $p(a)\neq p(b)$ changes the nature of the color
state from bosonic to fermionic or vice versa.

One can construct from these matrix units a representation of
the symmetric group ${\bf S}_N$. For this purpose, consider $N$
copies of the matrix unit $X_j^{ab}\;,j=1,\cdots,N$
that act on the graded vector space $V_1\otimes\cdots\otimes V_N$
where the subscript $j$ corresponds to the space $V_j\cong V$
in the tensor product. With the relation
\begin{equation}
X^{ab}_jX^{cd}_k=(-1)^{(p(a)+p(b))(p(c)+p(d))}X_k^{cd}X_j^{ab}\;,
\end{equation}
one can show that the operator defined as
\begin{equation}
P_{jk}=\sum_{a,b=1}^{n+m}(-1)^{p(b)}X_j^{ab}X_{k}^{ba}\label{eq:exc}
\end{equation}
exchanges the basis vectors $v_j$, $v_k$ of the $j,k$ space as
\begin{equation}
P_{jk}v_j^{a}\otimes v_k^{b}=(-1)^{p(a)p(b)}v_j^{b}\otimes v_k^{a}\;.
\end{equation}
and satisfies the relations
\begin{equation}
P_{ij}=P_{ji},\hspace{1cm}P_{ij}P_{ij}=1,\hspace{1cm}{\rm and}\hspace{1cm}
P_{ij}P_{jk}=P_{ik}P_{ij}\;.
\end{equation}
Hence it furnishes a representation for the symmetric group ${\bf S}_{N}$

Using this supersymmetric representation for the
symmetric group it is straight forward to construct the Yangian deformation
of $Y(gl(n|m))$. The $R_{oo'}(u)$ matrix acting on $V_o\otimes V_{o'}$
has the same expression as (\ref{eq:rmx}), except that the exchange
operator $P_{oo'}$ has a representation given in (~\ref{eq:exc})
with $j,k$ replaced by $o,o'$ that denote the auxiliary spaces
respectively. Similarly let
$\{T_s^{ab}\;,s\geq 0\;, a,b=1,\cdots,n+m\}$ be the Yangian $Y(gl(n|m))$
generators and define the $T$-matrix as
\begin{equation}
T_o(u)=\sum_{a,b=1}^{n}X_o^{ba}(-1)^{p(a)}T^{ab}(u)
\end{equation}
where
\begin{equation}
T^{ab}(u)\equiv {\bf 1}\delta_{ab}(-1)^{p(a)}
+\lambda\sum_{s\geq 0}u^{-s-1}T_s^{ab}\;.\label{eq:tmx}
\end{equation}
The Yangian relation is then encoded in (~\ref{eq:yan}). This defining
relation written in terms of $T_s^{ab}$ is given by
\begin{equation}
\left[T_s^{dc}\right.,\left.T_{p+1}^{ba}\right\}
-\left[T_{p+1}^{dc}\right.,\left.T_s^{ba}\right\}
=\lambda(-1)^{(p(b)p(d)+p(c)p(b)+p(c)p(d))}\left(T_p^{bc}T_s^{da}-T_s^{bc}T_p^{da}\right)
\label{eq:def}
\end{equation}
for $s,p\geq -1$  where $T_{-1}^{ab}\equiv {\lambda}^{-1}{\bf 1}\delta_{ab}$
and
the graded bracket here has the same definition as that in (\ref{eq:lieb}).

The center of the Yangian algebra can likewise be expressed in a
supersymmetric generalized quantum determinant\cite{naz}. However,
for our purpose we employ an alternative expression for the generator
of the center elements. Define the $\tilde{T}^{ab}_s$ operators
by the following condition\cite{naz};
\begin{equation}
\sum_{b}(-1)^{p(b)}T^{ab}(u)\tilde{T}^{bc}(u)=(-1)^{p(a)}\delta_{ac}\;,
\end{equation}
where $\tilde{T}^{ab}(u)$ is defined in terms of $\tilde{T}^{ab}_s$
as in (\ref{eq:tmx}).
This gives, for example, the first few modes of $\tilde{T}^{bc}(u)$
\begin{equation}
\begin{array}{lcl}
\tilde{T}_0^{bc}&=&-T_0^{bc}\\
\tilde{T}_1^{bc}&=&-T_1^{bc}+\lambda\sum_f(-1)^{p(f)}T_0^{bf}T_0^{fc}\\
\tilde{T}_2^{bc}&=&-T_2^{bc}+\lambda\sum_f(-1)^{p(f)}(T_0^{bf}T_1^{fc}
+T_1^{bf}T_0^{fc})\\
&&-\lambda^2\sum_{f,g}(-1)^{p(f)+p(g)}T_0^{bf}T_0^{fg}T_0^{gc}
\end{array}\;.
\end{equation}

The generator $Z(u)$ of the center is then given by
\begin{equation}
Z(u)=\left\{\begin{array}{ll}
\sum_{a,b}(-1)^{p(b)}T^{ab}(u+(n-m)\lambda)\tilde{T}^{ba}(u)/(n-m)&
\;;n\neq m\\ &\\
1+\sum_{a,b}(-1)^{p(b)}\frac{d}{du}(T^{ab}(u))\tilde{T}^{ba}(u)&\;;n=m\\
\end{array}\right.
\end{equation}

Expanding $Z(u)$ in power series of $u^{-1}$, we get
\begin{equation}
Z(u)={\bf 1}+\lambda u^{-2}Z_0+2\lambda u^{-3}Z_1+3\lambda u^{-4}Z_2+\cdots
\end{equation}
where
\begin{eqnarray*}
Z_0&=&Q_0\\
Z_1&=&Q_1+\frac{1}{2}(n-m)\lambda Q_0\\
Z_2&=&Q_2+(n-m)\lambda Q_1+\frac{1}{3}(n-m)^2\lambda^2 Q_0
\end{eqnarray*}
with
\begin{equation}
\begin{array}{lll}
Q_0&=&\sum_{a}T_0^{aa}\\
&&\\
Q_1&=&\sum_{a}T_1^{aa}-\frac{1}{2}\lambda\sum_{a,b}(-1)^{p(b)}T_0^{ab}T^{ba}_0\\
&&\\
Q_2&=&\sum_{a}T_2^{aa}-\lambda\sum_{a,b}(-1)^{p(b)}T_0^{ab}T^{ba}_1
+\frac{1}{3}\lambda^2\sum_{a,b,c}(-1)^{(p(b)+p(c))}T_0^{ab}T^{bc}_0T^{ca}_0\\
&&\mbox{}-\frac{1}{6}\lambda(n-m)\sum_{a,b}(-1)^{p(b)}T_0^{ab}T_0^{ba}\;,
\end{array}\label{eq:central}
\end{equation}
which are the first few central elements of interest to us.
Notice that unlike the mode expansion of the quantum determinant, here
the coefficient of $u^{-1}$ vanishes.

It is clear that the special case $n=0(m=0)$ corresponds to the Yangian
$Y(gl(n))$ with deformation parameter $\lambda(-\lambda)$. From (\ref{eq:def})
we see that interchanging the grading of the basis vector to
\[p(a)=\left\{\begin{array}{ll}
           1&\hspace{2cm}a=1,\cdots,n\\
           0&\hspace{2cm}a=n+1,\cdots,n+m
\end{array}\right.\]
corresponds to changing $\lambda$ to $-\lambda$. Physically this will
correspond to interchanging the bosonic and fermionic color.

In the next section we will study a particular realization of
this supersymmetric Yangian which has physical interest.

\section{$gl(n|m)$ color Calogero-Sutherland model}
Consider a collection of $N$ identical particles described by
the set of coordinates $\{x_j\}$ and carry color index $a=1,\cdots,n+m$.
By definition, the particles are called bosons (fermions)
if the wavefunction is symmetric (anti-symmetric) under
exchange of the particles (i.e. coordinates and color indices).
As mentioned in previous section, depending on the value of the color index,
the color degrees of freedom can have a bosonic or fermionic character
according to the corresponding ${\bf Z}_2$ grading, in order not to
confuse this fermionic/bosonic nature of the {\em color} with
fermionic/bosonic identity of the {\em particle} (as defined by the
property of the wave function under exchange operation),
we shall always refer to the former as fermionic/bosonic-color.
For definiteness, restrict to the case of $N$ identical bosons,
the configuration space is given by $\mbox{\bf Sym}(V^{\otimes N}\otimes
{\bf C}[x_1,\cdots,x_N])$ where
${\bf C}[x_1,\cdots,x_N]$ denotes the set of complex functions in the
variable $\{x_1,\cdots,x_N\}$ and {\bf Sym} the projection onto the subspace
which is symmetric under simultaneous exchange of $x_j,x_k$ and $v_j, v_k$
for any $j,k$. Note that such restriction is justified as long as the
the Hamiltonian is invariant under this operation. In general, for a
mixture of fermions and bosons, one has to consider projection onto
the subspace which corresponds to the Young tableau that describes
the exchange properties of the collection of the particles.

 From the configuration space, one can construct a representation of
the supersymmetric Yangian. To start, we define the action of
$T_0^{ab}$ and $T_1^{ab}$ on this space as follows:
\begin{eqnarray}
T^{ab}_0&=&\sum_{j=1}^N X^{ab}_j  \\
T^{ab}_1&=&\sum_{j,k=1}^N X^{ab}_jL_{jk}
\end{eqnarray}
where
\[L_{jk}=\delta_{jk}\frac{\partial}{\partial x_j}+\lambda(1-\delta_{jk})
\theta_{jk}P_{jk}\]
is an operator that plays the role of one of the Lax pair element
in \cite{wat}, and $X^{ab}_j,P_{jk}$ have been defined in the previous section,
while the function $\theta_{jk}\equiv\theta(x_j-x_k)$ is
yet to be determined.

Next, we require that these $0,1$ modes of the $T$-matrix
satisfy the Yangian relation (~\ref{eq:def}). It is easy to check
that (\ref{eq:def}) hold for $s,p=-1,0$. For other values of $s,p$,
the Yangian relation can be considered as a relation that generates
higher mode of the $T$-matrix from the lower modes ones. Requiring that
the $T$-matrix generated be consistent imposes conditions on the
function $\theta_{jk}$. Consider, for example, $s=1, p=0$ in (\ref{eq:def}),
we get
\begin{eqnarray}
\left[T_1^{ba}\right.,\left.T_{1}^{dc}\right\}&
=&\delta_{ad}T_{2}^{bc}-
(-1)^{(p(a)+p(b))(p(c)+p(d))}T_2^{da}
\nonumber\\
&&\mbox{}+\lambda(-1)^{(cd+a(c+d))}\left(T_0^{bc}T_1^{da}-
T_1^{bc}T_0^{da}\right)\;,
\label{eq:def1}
\end{eqnarray}
substituting the representation of $T_0^{ab},T_1^{ab}$ into the above,
we find that the function $\theta_{jk}$ satisfies exactly the same
relations given in (\ref{eq:fre}), the general solution is therefore
given by
\begin{equation}
\theta_{jk}=(1-a^{x_j-x_k})^{-1}\;,
\label{eq:tjk}
\end{equation}
where $a$ is an arbitrary ${\bf C}$ number. In addition, the above
also gives
\begin{equation}
T_2^{ab}=\sum_{j,k=1}^{N}X^{ab}_j (L^2)_{jk}\;.
\end{equation}
Similarly, we find
\begin{equation}
T_3^{ab}=\sum_{j,k=1}^{N}X^{ab}_j (L^3)_{jk}
\end{equation}
and one can check that (\ref{eq:fre}) is the necessary condition
for $T_3^{ab}$ to be generated consistently.
In an alternative definition of the Yangian algebra\cite{cha}
only $\{T_{s}^{ab};\ s=0,1,2,3\}$ are needed. Hence, it suffices
to check the above construction up to $T_3^{ab}$.

It is no coincidence that the function $\theta_{jk}$ satisfies the
same relation as that of the CS model. To see this, note that since
the wavefunction is invariant under exchange of two particles, one can
trade the colors exchange operator $P_{jk}$ with the coordinates exchange
operator $K_{jk}$ and following the approach of \cite{sac} to
show that
\begin{equation}
T^{ab}(u)={\bf 1}\delta_{ab}+\sum_{j=1}^{N}\frac{X_j^{ab}}{u-D_j}
\end{equation}
with
\[D_j=\frac{\partial}{\partial x_j}+\lambda\sum_{k\neq j}\theta_{jk}K_{jk}\]
satisfies the Yangian defining relation. Furthermore, this also
implies that
\begin{equation}
T^{ab}_s=\sum_{j,k}X^{ab}_j(L^s)_{jk}\;;\hspace{1.5cm}s\geq 0
\label{eq:gen}\;.
\end{equation}
The same argument can also be applied to the case when all
particles are fermions or even mixture of fermions and bosons.
In which case the $P_{jk}$ is replaced with $\pm K_{jk}$ depending
on the bosonic/fermionic identities of particles $j,k$, and the
proof used in \cite{sac} still hold.

With the representation for $T_s^{ab}$ we can compute the central
elements of the supersymmetric Yangian. The first few
central elements in (\ref{eq:central}) become
\begin{eqnarray}
Q_0&=&N\\
Q_1&=&P-\frac{1}{2}\lambda N(n-m)\\
Q_2&=&2 H-\lambda(n-m)P-\frac{1}{6}\lambda N(N-1)+\frac{1}{6}\lambda^2(n-m)^2N
\end{eqnarray}
where $P,H$ are the total momentum and Hamiltonian of the system
of particles given by
\begin{eqnarray}
P&=&\sum_{j}\frac{\partial}{\partial x_j}\\
H&=&\frac{1}{2}\sum_{j,k}\left(\partial_j^2+\lambda P_{jk}\partial_j\theta_{jk}
+\lambda^2 \theta_{jk}\theta_{kj}\right)\;.
\label{eq:hamil}
\end{eqnarray}

So from this construction, it is clear that the rest of the
central elements, which commute with $H$, are conserved quantities of the
model and they commute among themselves, so the model is integrable.
Using (\ref{eq:tjk}), the Hamiltonian can be written explicitly
\begin{equation}
H=\frac{1}{2}\sum_{j}\frac{\partial^2}{\partial x_j^2}
+\sum_{j>k}\frac{\lambda(\lambda-P_{jk})}{\sinh^2[{a\over{2}}(x_j-x_k)]}\;.
\end{equation}

Notice indeed that the Hamiltonian is invariant under simultaneous
permutation of particles' coordinates and color indices, hence
the justification of restricting the Hilbert space to be the subspace of
$V^{\otimes N}\otimes {\bf C}[x_1,\cdots,x_N]$ which are irreducible
module of the symmetric group ${\bf S}_N$. As $T^{ab}_s$ leave the
Hilbert space invariant, the action of, say, $T^{ab}_sT^{ab}_p$ on
a totally symmetric wavefunction (ie. $N$ identical Bosons case) to remain
symmetric even though for $p(a)\neq p(b)$ some of the color degrees of freedom
will be changed from bosonic to fermionic or vice versa. On the other hand,
if two bosonic particles $j,k$ have fermionic colors in the above
symmetric wavefunction, letting the quantum numbers of these two particles
to be equal, we see that the wave function has to vanish.

\section{Special  limits of the SUSY CS model}

First we study the rational limit of the model.
This can be achieved by
rescaling $\lambda=a\lambda^{'}$ and letting $a\rightarrow 0$.
The function $\theta_{jk}$ becomes $\omega_{jk}$ defined as
\begin{equation}
\omega_{jk}=\frac{1}{x_j-x_k}\;,
\end{equation}
while the Hamiltonian becomes
\begin{equation}
H=\frac{1}{2}\sum_{j,k}\left(\partial_j^2+\lambda^{'} P_{jk}\partial_j
\omega_{jk}+\lambda^{'2} \omega_{jk}\omega_{kj}\right)\;.
\end{equation}
The $L_{jk}$ operator remains non-vanishing as
\begin{equation}
L_{jk}=\delta_{jk}\frac{\partial}{\partial x_j}
+\lambda^{'}\omega_{jk}P_{jk}
\end{equation}
but the Yangian relation, under this rescaling becomes
\begin{equation}
\left[T_m^{dc}\right.,\left.T_{n+1}^{ba}\right\}
-\left[T_{m+1}^{dc}\right.,\left.T_n^{ba}\right\}=0
\end{equation}
with $T_{-1}^{ab}\equiv \delta_{ab}{\bf 1}$. The above can be
recasted into the more familiar form given  by
\begin{equation}
\left[T_n^{ab}\right.,\left.T_m^{cd}\right\}
=\delta_{bc}T_{n+m}^{ad}-(-1)^{(p(a)+p(b))(p(c)+p(d))}\delta_{ad}T_{n+m}^{cb}\;.\label{eq:loop}
\end{equation}
Hence the symmetry becomes the $gl(n|m)$ loop algebra. The
central elements are simply given by
\begin{equation}
Q_s=\sum_{a}T_s^{aa} \hspace{2cm}s\geq0\;,
\end{equation}
which for the case $n=m$, satisfy
\begin{equation}
Q_{s+p}={\sum_{a,b}}'\left\{T^{ab}_s,T_p^{ba}\right\}
\end{equation}
where $\{\;,\;\}$ is the usual anti-commutator and $\sum^{'}$ denotes
summation restricted to  color indices with $p(a)=0,p(b)=1$. These generators
have grading 1 and satisfy
\[\left\{T^{ab}_s,T_p^{cd}\right\}=0\hspace{1cm}\mbox{for }
p(a)=p(c)\neq p(b)=p(d)\;.\]
Therefore they can be regarded as supersymmetric charges. In particular,
the Hamiltonian, which is simply $\frac{1}{2}Q_2$, is written as
\[H=\frac{1}{2}{\sum_{a,b}}^{'}\left\{T^{ab}_1,T_1^{ba}\right\}\;.\]

In this light, we can consider the $Y(gl(n|m))$ Yangian algebra as
a deformation of the SUSY algebra given above.
Only in the rational limit, the $gl(n|m)$ CS model can have the supersymmetry.

Restricting further to the case where the symmetry is the
$gl(1|1)$ loop algebra, the model is the supersymmetric CS model studied in
\cite{men} without the harmonic potential. To see this, consider $N$
copies of Grassmanian variables and their conjugates
$\{\theta_j,\theta_j^{\dagger}\}$, they satisfy the usual anti-commutation
relations
\begin{equation}
\{\theta_j,\theta_k\}=\{\theta^{\dagger}_j,\theta^{\dagger}_k\}=0
\hspace{1cm}\{\theta_j,\theta^{\dagger}_k\}=\delta_{jk}\;.
\end{equation}
For a given $j$, the set $\{\theta_j, \theta_j^{\dagger}\}$  furnishes a
two dimensional representation for the $gl(1|1)$ as
\begin{equation}
\begin{array}{lcllcl}
X_j^{12}&=&\theta_j^{\dagger}&X_j^{21}&=&\theta_j\\
X_j^{11}&=&\theta_j^{\dagger} \theta_j&X_j^{22}&=&\theta_j
\theta_j^{\dagger}\;.
\end{array}
\end{equation}
The $T_1^{ab}$ generator given by
\begin{equation}
T_1^{ab}=\sum_{j}X_j^{ab}\frac{\partial}{\partial x_j}+\lambda^{'}
\sum_{j\neq k}
\omega_{jk}(X_j^{aa}X_k^{ab}(-1)^{p(a)}
+X_j^{ab}X_k^{bb}(-1)^{p(b)})\hspace{1cm} a\neq b
\end{equation}
can be simplified by noting that the last term can be combined into
$\lambda^{'}\omega_{jk}(-1)^{p(a)}X_{j}^{ab}$. Upon substituting the
Grassmanian variables for $X_j^{ab}$, we get
\begin{eqnarray}
T_1^{12}&=&\sum_{j}\theta^{\dagger}_j\left(\frac{\partial}{\partial x_j}
+\lambda^{'}\sum_{k\neq j}\omega_{jk}\right)\\
T_1^{21}&=&\sum_{j}\theta_j\left(\frac{\partial}{\partial x_j}
-\lambda^{'}\sum_{k\neq j}\omega_{jk}\right)\;,
\end{eqnarray}
which coincide with the SUSY
charges $Q^{\dagger},Q$ in \cite{men}.

Another limit which is of interest is given by $\lambda\rightarrow\infty$,
where the $L$-operator takes the simplified form
\[L_{jk}=\lambda\theta_{jk}P_{jk}\;.\]
Since $T_s^{ab}\sim \lambda^s$, the Yangian relation (\ref{eq:yan}) actualy
does not depend on $\lambda$ and remains finite in this limit. Thus
$T^{ab}_s$'s constructed with the above $L$-operator give a
representation of the Yangian $Y(gl(n|m))$. However, one can check that with
this representation, the generator of the central elements $Z(u)$
is proportional to the identity operator, this can be easily
seen in $Q_0,Q_1, Q_2$ keeping only the most dominant terms.
On the other hand, consider the expression for $H$ in (\ref{eq:hamil}).
It can be shown that the second term given by
\begin{equation}
\frac{1}{2}\lambda\sum_{j\neq k} \partial_{j}\theta_{jk}P_{jk}
\end{equation}
continues to commute with the Yangian generators. One can then
take the above as the Hamiltonian of a spin system, where spins here
refer to the elements of the vector module $V^{\otimes N}$ of
$gl(n|m)$ and the spins are fixed at equilibrium positions of the
CS particles. This is therefore a supersymmetric extension of the
Haldane-Shastry spin chain. The motivation for ignoring the
most dominant term in $H$ is that it is an unimportant constant,
it is not clear whether one can also obtain other commuting
conserved quantities from central elements of the origin Yangian
$Y(gl(n|m))$ following the above procedure. Nonetheless, one can employ
the approach used in \cite{poly} to construct other commuting conserved
quantities and show that this spin system is integrable.

On the other hand, it is well-known\cite{sac} that the $SU(2)$ Haldane-Shastry
model corresponds to CS model without color for a special value of $\lambda$.
In our case, we expect a similar correspodence appears so that
$gl(1|1)$ CS model may be related to the spin chain model with $gl(1|2)$
color. Furthermore, it is likely that one may find relevance of this
spin chain model with the supersymmetric $t-J$ model considered in
\cite{kuramoto}.

\section{Conclusion}

The present paper gives a unified treatment for the color $gl(n|m)$
Calogero-Sutherland models. We prove that the family of models is
integrable with the Yangian $Y(gl(n|m))$ as the underlying symmetry.
 This family of the models include the well studied $gl(n)$ extension of
CS model. In the rational limit, we demonstrate how the supersymmetry
charges arise based on the $gl(n|m)$ loop algebra.

The work is a modest effort in trying to understand the
underlying algebraic structure of the integrable long range interacting
models. By identifying the common feature, one may hope to
extend the construction to models related to other Lie algebra as in
the case of integrable nearest-neighbor interacting spin systems.
In fact there are more general CS type models associated with all
root systems of the simple Lie algebra\cite{ols}, it would be
interesting to identify the symmetry of these models.

One can also relate the long range nonrelativistic model
to the factorizable scattering theory without implementing the crossing
symmetry\cite{aln}. Here, we claim that the supersymmetric factorizable
$S$-matrix\cite{ShaWit} in the nonrelativistic limit can be obtained
from the two-particle wave functions of the supersymmetric CS model
with $1/\sinh^2(x)$ potential\cite{next}.

\section*{Acknowledgment}
We thank C. Lee for useful discussion. W.M.K acknowledges stimulating
conversation with M.L.Ge.
CA is supported in part by KOSEF 941-0200-003-2 and BSRI 94-2427 and
WMK by a grant from the
Ministry of education and KOSEF through SNU/CTP.

\end{document}